\documentclass[aps,prc,twocolumn,showpacs,nofootinbib]{revtex4}%
\usepackage{amsfonts}
\usepackage{amsmath}
\usepackage{amssymb}
\usepackage{graphicx}
\usepackage{dcolumn}%
\def\bea{\begin{eqnarray}}
\def\eea{\end{eqnarray}}

\begin{document}
\title{Goldberger-Treiman constraint criterion for hyperon coupling
constants}
\author{Ignacio J. General}
\affiliation{Department of Physics, North Carolina State University,
Raleigh, North
Carolina 27695-8202}
\author{Stephen R. Cotanch}
\affiliation{Department of Physics, North Carolina State University,
Raleigh, North
Carolina 27695-8202}
\keywords{coupling constant, pion, kaon, Goldberger-Treiman Relation,
Goldberger-Treiman
Discrepancy, Dashen-Weinstein sum rule}
\pacs{11.30.Hv, 11.40.Ha, 13.30.Ce, 13.75.Gx, 13.75.Jz}

\begin{abstract}
The generalized Goldberger-Treiman relation is combined with the
Dashen-Weinstein sum rule to provide a constraint equation between
the $g_{K\Sigma N}$ and $g_{K\Lambda N}$ coupling constants. A
comprehensive examination of the published phenomenological and
theoretical hyperon couplings has yielded a much smaller set of
values, spanning the intervals $0.80\leq g_{K\Sigma
N}/\sqrt{4\pi}\leq2.72$ and $-3.90\leq g_{K\Lambda
N}\mathbf{/}\sqrt{4\pi}\leq-1.84$, consistent with this criterion.
The   $SU_F(3)$ and Goldberger-Treiman hyperon couplings satisfy
the constraint along with predictions from a Taylor series
extrapolation using the same momentum variation as exhibited by
$g_{\pi NN}$.
\end{abstract}
\volumeyear{year}
\volumenumber{number}
\issuenumber{number}
\eid{identifier}
\date{\today}
\startpage{1}
\endpage{2}
\maketitle







\section{Introduction}

After a half century of investigating meson-baryon interactions,
it is somewhat surprising that there are still several important
coupling constants not accurately known.  While the $\pi N$
coupling constant, $g_{\pi N N}$, has been determined to within a
few percent, significant uncertainty in the two hyperon couplings,
$g_{K(Y=\Lambda, \Sigma)N}$, remains and even recently published
values for both vary by more than a factor of four. This large
variance is due to limited experimental information and also
shortcomings in theoretical models.  Further, analyses of purely
hadronic processes typically yield larger couplings than those
obtained from hyperon electromagnetic production studies.
Fortunately, with the advent of new accelerator facilities, such
as Jefferson Lab and SPring-8, more accurate and abundant data are
now becoming available.  Related, the recently
reported~\cite{theta} discovery of the exotic strangeness +1
pentaquark resonance, $\Theta^+$, is also attracting attention
which should spawn additional $KN$ measurements.

The purpose of this work is to detail a potentially useful
constraint relation between $g_{K \Lambda N}$ and $g_{K \Sigma N}$
which should facilitate future hyperon scattering and production
analyses, especially with respect to extracting more accurate
coupling constants.  The constraint involves the generalized
Goldberger-Tremain (GT) relation~\cite{GT} and the
Dashen-Weinstein (DW) sum-rule~\cite{Dashen-Weinstein}. The GT
relation is exact in the combined chiral and zero momentum limits
according to the partially conserved axial vector current (PCAC)
hypothesis and the assumed slow momentum variation of the $\pi N$
coupling constant. Even with explicit chiral symmetry breaking,
the nucleon GT relation remains valid and is now satisfied to
within one percent (see section II). Because of the larger strange
quark mass and attending broken $SU_F(3)$ flavor symmetry, the
generalized GT relation is not as accurate in the hyperon sector.
However the deviation, or hyperon GT discrepancy, $\Delta_Y$, is
believed to be reasonably accurately constrained by the DW sum
rule since corrections are suppressed by two powers in the heavy
baryon chiral perturbation theory expansion~\cite{Goity}.
Accordingly, by utilizing the DW sum rule connecting the $N$,
$\Lambda$ and $\Sigma$ discrepancies, we have obtained a
presumably accurate constraint equation between the hyperon and
nucleon coupling constants, involving the known axial charges and
hadron masses.

There have been several studies~\cite{Goity,Dominguez,bh,Fuchs}
that have implemented the DW sum rule, especially to constrain the
$\pi N$ coupling constant~\cite{Goity,Dominguez,bh}. Our approach
represents a different view as we submit the hyperon couplings and
their discrepancies are the limiting, less accurate quantities and
that the reasonably well known $g_{\pi N N}$, through the DW sum
rule, provides a constraint for $g_{K \Lambda N}$ and $g_{K \Sigma
N}$.  Because $g_{\pi N N}$ is only determined to within a few
percent, the constraint equation produces a band in the ($g_{K
\Sigma N}$, $g_{K \Lambda N}$) plane and we document which
published couplings, when plotted, fall within this band. Since
there is only one equation for the two couplings our constraint
will only be useful for analyses involving both couplings.
However, this should encompass most phenomenological
investigations since models for hyperon reactions and production
entail both $\Lambda$ and $\Sigma$ intermediate states and their
attending couplings.  Because of this interdependence our
criterion should be useful even if analyses of purely hadronic
processes continue to provide larger hyperon couplings than
electromagnetic production (i.e. the constraint should provide a
good numerical relation between the two couplings even if there is
an effective coupling renormalization due to model dependence).
Our result should also be of special interest to the hyperon and
hypernuclear community and of timely benefit in the analysis of
precision kaon electromagnetic production data recently
measured~\cite{schum} at Jefferson Lab. Obtaining improved hyperon
coupling constants will also permit new confrontations with QCD
based
theoretical approaches which have been successful in calculating $g_{\pi NN}%
$~\cite{ep}.

This paper is organized into four sections. The generalized GT
relation and the DW sum rule, along with the $SU_{F}(3)$ coupling
relations, are given in section II and the constraint criterion is
developed. Published values for $g_{KYN}$ are reviewed in section
III and the coupling constraint is imposed producing a subset
satisfying this criterion. Finally, conclusions are summarized in
section IV.

\section{Fundamental Coupling relations}

\subsection{Generalized Goldberger-Treiman relation}

To fully appreciate the validity of the GT relation it is illustrative to
sketch its
derivation. There are several ways to obtain this result such as using PCAC or
unsubtracted dispersion relations (pion pole dominance of the axial vector
divergence). Here the PCAC approach is adopted. Consider the matrix element of
the axial $SU_{F}(3)$ current operator, $j_{a}^{\mu5}(x)=\overline{\Psi
}\left(  x\right)  \gamma_{\mu}\gamma_{5}\frac{\lambda^{a}}{2}\Psi\left(
x\right)  $, between two baryon octet states, $\left\langle
B\right\vert j_{a}^{\mu5}\left\vert B^{\prime}\right\rangle $. Since the
current transforms as a pseudovector, the most general form for this matrix
element is
\bea
\left\langle B\right\vert j_{a}^{\mu5}(x)\left\vert B^{\prime}\right\rangle
= e^{  -i \ q
\cdot x } \overline{u}\left(  p\right)
  [  g^B_{A}\left(  q^{2}\right)  \gamma^{\mu}\gamma_{5} \nonumber
\\ + \ g^B_{T}\left(
q^{2}\right)  i\sigma^{\mu\nu}q_{\nu}\gamma_{5}+g^B_{P}\left(  q^{2}\right)
q^{\mu}\gamma_{5} ]  u\left(  p^{\prime}\right)   ,
\label{ncurrent}
\eea
where $q=p-p^{\prime}$, ${p}^{2}=m^2_{B}$, ${p^{\prime}}^{2}=m^2_{B^{\prime
}}$ and $g_{A}^{B}\left(  q^{2}\right)  $ and $g_{P}^{B}\left(
q^{2}\right)  $ are the axial vector and induced pseudoscalar form factors,
respectively. The induced tensor form factor, $g_{T}^{B}\left(  q^{2}\right)
$, violates $G$-parity and will be omitted, consistent with small effects from
second class currents. The axial current operator also appears in the
definition of the decay constant, $f_{M}$, for a pseudoscalar octet meson
$M^{b}$ having mass $m_{M}$
\begin{equation}
\left\langle 0\right\vert j_{a}^{\mu5}\left(  x\right)  \left\vert
M^{b}\left(  q\right)  \right\rangle =i\sqrt{2}f_{M}q^{\mu}\delta_{a}%
^{b}e^{-i\ q\cdot x}. \label{4}%
\end{equation}
Taking the divergence
\begin{equation}
\left\langle 0\right\vert \nabla_{\mu}j_{a}^{\mu5}\left(  x\right)
\left\vert M^{b}\left(  q\right)  \right\rangle =\sqrt{2}f_{M}m_{M}^{2}%
\delta_{a}^{b}e^{-i\ q\cdot x},
\end{equation}
yields a conserved axial current in the generalized chiral limit ($\lim_{m_{M}%
\rightarrow0}\nabla_{\mu}j_{a}^{\mu5}=0$); this is the PCAC hypothesis.
Under this assumption the baryon axial vector current, Eq. (\ref{ncurrent}),
is also conserved, and its divergence yields
\begin{equation}
\overline{u}\left(  p\right)  \left[  g_{A}^{B}\left(  q^{2}\right)
/\!\!\!q+g_{P}^{B}\ q^{2}\right]  \gamma_{5}u\left(  p^{\prime}\right)  =0\ .
\end{equation}
Then using the free Dirac equation for the first term leads to the form factor
relation
\begin{equation}
g_{A}^{B}(q^{2})=-\frac{q^{2}}{m_{B}+m_{B^{\prime}}} \ g_{P}^{B}(q^{2}) \ .
\label{g1}
\end{equation}
To proceed further, consider the leading Feynman diagrams for the weak
decay $B\rightarrow
B^{\prime}+\overline{\nu}_{l}+l$ depicted in Fig. 1.
\begin{figure}[h]
\begin{center}
\includegraphics[height=2.5131in,width=3.0381in]{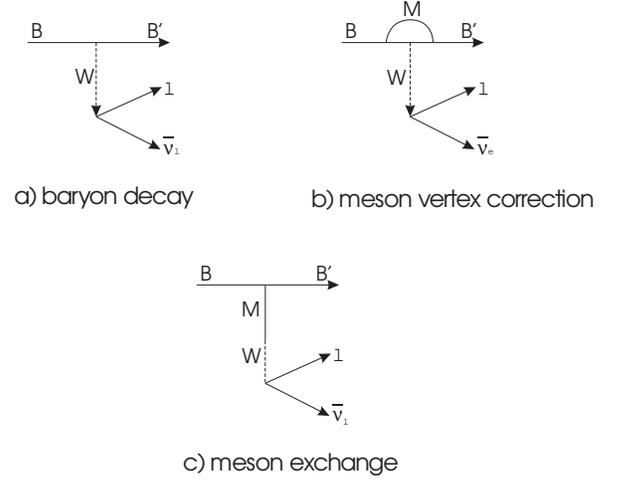}
\caption{\label{decays}Baryon weak decay and meson corrections.}
\end{center}
\end{figure}
Only the meson exchange graph c)  contributes to $g_{P}^{B}$.
Direct evaluation gives
\begin{equation}
g_{P}^{B}\left(  q^{2}\right)  =-\frac{\sqrt{2}f_{M}}%
{q^{2}-m_{M}^{2}} g_{MBB^{\prime}} \ , \label{g3}%
\end{equation}
where $g_{MBB^{\prime}}$\ is the strong interaction baryon-meson coupling
constant. Including higher order vertex corrections would modify this result
by an additional multiplicative form factor, $F(q^{2})$, with $F(0)=1$.
Combining Eqs. (\ref{g1},\ref{g3}) finally yields in the combined chiral and
zero momentum limits
\begin{equation}
g_{A}^{B}(0)=\frac{\sqrt{2}f_{M}}{m_{B}+m_{B^{\prime}}}g_{MBB^{\prime}}(0) \ ,
\end{equation}
or rearranging
\begin{equation}
g^{GT}_{MBB^{\prime}} \equiv g_{MBB^{\prime}}(0) = \frac{
(m_{B}+m_{B^{\prime}})}
{\sqrt{2}f_{M}}g_{A}^{B}(0) \ .
\label{ggtr}
\end{equation}
This is the generalized GT relation that defines the GT coupling
constant and is exact in the zero meson mass and momentum limits,
$m_{M}^{2}=q^{2}=0$.

We now apply Eq. (\ref{ggtr}) to evaluate the GT coupling constants $g_{\pi
NN}^{GT}$ and $g_{KYN}^{GT}$. First we specify our coupling constant
convention and phase consistent with the usual pseudoscalar Lagrangian
\bea
\mathcal{L}=ig_{\pi NN}
\overline{N}\gamma_{5}{\mbox {\boldmath${\tau}$\unboldmath}}N
\cdot{\mbox {\boldmath${\pi}$\unboldmath}}+ig_{K\Lambda N}
\overline{N}\gamma_{5}
\Lambda K\nonumber
\\ + ig_{K\Sigma N}  \overline{N}\gamma_{5}
{\mbox{\boldmath${\tau}$\unboldmath}}\cdot\mbox{{
\boldmath${\Sigma}$\unboldmath}}K + h. c.
\ ,
\eea
with isospin nucleon, $N={\binom{p}{n}}$, and kaon, $K={\binom
{K^{+}}{K^{0}}}$, doublet,  and   pion, {\mbox
{\boldmath${\pi}$\unboldmath}}, and sigma,
{\mbox
{\boldmath${\Sigma}$\unboldmath}}, triplet fields. The different meson
charge couplings are
related to the generic coupling constants by
\begin{align}
g_{\pi NN}  &  \equiv g_{\pi^{0}nn}=-\frac{1}{\sqrt{2}}g_{\pi^{-}np}\ ,\\
g_{K\Lambda N}  &  \equiv g_{K^{-}\Lambda p}\ ,\\
g_{K\Sigma N}  &  \equiv\frac{1}{\sqrt{2}}g_{K^{-}\Sigma^{-}n}\ .
\end{align}
With this notation and the most recently measured parameters~\cite{PDG} listed
in Table~\ref{tab:model}, including the axial charges (note
$g^N_{A}=-g_{A}^{n}%
$) corresponding to the weak decays $n\rightarrow p+e^{-}+\overline{\nu}_{e}$,
$\Lambda\rightarrow p+e^{-}+\overline{\nu}_{e}$ and $\Sigma^{-}\rightarrow
n+e^{-}+\overline{\nu}_{e}$, the GT coupling constants are
\begin{align}
g_{\pi NN}^{GT}  &  =\frac{g^N_{A}%
}{f_{\pi}}\frac{(m_{n}+m_{p})}{2}=12.897 \pm 0.047\ ,\label{gngt}\\
g_{K\Lambda N}^{GT}  &  =\frac{g_{A}^{\Lambda}}{\sqrt{2}f_{K}}\left(
m_{\Lambda}+m_{p}\right)  =-9.228 \pm 0.209\ ,\label{glgt}\\
g_{K\Sigma N}^{GT}  &  =\frac{g_{A}^{\Sigma}}{f_{K}}\frac{\left(  m_{\Sigma
}+m_{n}\right)  }{2}=3.215 \pm 0.163\ . \label{gsgt}%
\end{align}

Using these results and the commonly cited~\cite{de Swart 2} $\pi N$ coupling
 value $g_{\pi NN}=13.02 \pm 0.08$, it is interesting to make a simple
Taylor series extrapolation
for the hyperon couplings
\begin{equation}
g_{KYN}^{TS}= g_{KYN}(0) + m^2_{K} \frac{d g_{KYN}}{d q^2}(m^2_{K}) \ .
\end{equation}
Then assuming that the magnitudes of the GT coupling are lower bounds,
as suggested from results in section III, and using the same momentum
variation (derivative)
as exhibited by the $\pi N$ coupling, the predicted
Taylor series coupling constants are
\bea
|g_{KYN}^{TS}| &=& |g_{KYN}^{GT}| + m^2_{K} \frac{g_{\pi NN} - g_{\pi
NN}^{GT}}{m^2_{\pi}} \ , \\
    &=& |g_{KYN}^{GT}| + 0.123 \frac{m^2_{K}}{m^2_{\pi}} =
|g_{KYN}^{GT}| + 1.539 \ .
\nonumber
\eea
This yields $g_{K\Lambda N}^{TS} = -10.77 $ and $g_{K\Sigma N}^{TS} = 4.75
$ which will
also be assessed in section III along with the published hyperon couplings.

\begin{table}[h]
\caption{\label{tab:model}
Hadron masses, axial charges and decay constants. Errors are not listed for
the very accurately known masses.}
\begin{ruledtabular}
\begin{tabular}{l|r|l|c}
$m_{p} $ &  938.272 MeV  &   &  \\
$m_{n} $ &  939.565 MeV  & $g^N_A$  &  1.2695 $\pm$ 0.0029  \\
$m_{\Lambda}$ & 1115.683 MeV &  $g^{\Lambda}_A$  & -0.718 $\pm$ 0.015      \\
$m_{\Sigma^{-}}$ &  1197.449 MeV & $g^{\Sigma}_A$  &  0.340 $\pm$ 0.017
\\
$m_\pi$      &  139.570 MeV & $f_\pi$ & 92.42 $\pm$  .26 MeV   \\
$m_K$        &  493.677 MeV  & $f_K$        &  113.0  $\pm$ 1.0 MeV \\
\end{tabular}
\end{ruledtabular}%
\end{table}

\subsection{Dashen-Weinstein sum rule}

As stated above, the GT relation is exact in the combined chiral and zero
momentum limits.
The deviation of
$g_{MBB^{\prime}}^{GT}$ from the "physical"\footnote{There is  some
ambiguity as there is no single coupling constant because at least one
vertex particle is always off-shell and different processes will have
different particles off-shell.}, $g_{MBB^{\prime}}(m_{M}^{2})$, defines
the Goldberger-Treiman discrepancy (GTD)
\begin{equation}
\Delta_{B}\equiv1-\frac{g_{MBB^{\prime}}^{GT}}{g_{MBB^{\prime}}(m_{M}^{2})} \ .
\end{equation}

There are several relations for $\Delta_{B}$, one of which
is a sum rule first derived by Dashen and Weinstein \cite{Dashen-Weinstein}.
It connects the GTD for the
$\pi N$ and hyperon couplings and is given by
\begin{equation}
g_{\pi NN}\Delta_{N}=\frac{1}{2}\frac{m_{\pi}^{2}}{m_{K}^{2}%
}\left(  g_{K\Sigma N}\Delta_{\Sigma}-\sqrt{3}g_{K\Lambda N}\Delta_{\Lambda
}\right)  . \label{gapsum}%
\end{equation}
While this relation is an approximation, its validity appears to
be widely accepted~\cite{Dashen-Weinstein,Goity,Dominguez,bh,Fuchs} and, as
detailed in Ref. \cite{Goity},
corrections  are suppressed by two powers
in the heavy baryon chiral perturbation theory expansion.

From Eq. (\ref{gngt}) and the  $\pi N$ coupling $g_{\pi
NN}=13.02$, the nucleon discrepancy is $\Delta_N = .0094$
indicating that the GT relation is now satisfied to better than
1\%. Because $\Delta_{N}$ is much better known than
$\Delta_{Y=\Lambda ,\Sigma}$, we regard the DW sum rule as an
equation between  $\Delta_{\Lambda}$ and $\Delta_{\Sigma}$, which
permits extracting one of the coupling constants if the other is
known. Unfortunately neither is that accurately known so the sum
rule only provides a correlated constraint and this is the basis
of our hyperon coupling criterion. Rearranging Eq. (\ref{gapsum})
gives a linear constraint relation between $g_{K\Lambda N}$ and
$g_{K\Sigma N}$
\begin{equation}
g_{K\Lambda N}=\frac{1}{\sqrt{3}}g_{K\Sigma N}+b, \label{constraint}%
\end{equation}
where the intercept $b$ is given by
\begin{equation}
b= g_{K\Lambda N}^{GT}-\frac{g_{K\Sigma N}^{GT}}{\sqrt{3}} + \frac{2}{\sqrt
{3}}\frac{m_{K}^{2}}{m_{\pi}^{2}}[g_{\pi NN}^{GT}-g_{\pi NN}] \ . \label{b}%
\end{equation}
Because the small uncertainty in $g_{\pi NN}$ is magnified by the
large meson mass ratio, the constraint  only restricts an area in
the ($g_{K\Sigma N},g_{K\Lambda N}$) plane bounded by two parallel
lines with the maximum and minimum intercept values corresponding
to the error in $g_{\pi NN}$. Nevertheless, it still provides a
new criterion for evaluating the two $g_{KYN}$ couplings,
especially when they are analyzed in tandem. This coupling
constraint is applied to the published phenomenological and
theoretical $g_{KYN}$ in section III.

\subsection{SU(3) relation between coupling constants}

Unbroken $SU_{F}(3)$ flavor symmetry provides another relation between the
baryon-meson coupling constants~\cite{de Swart}. Using de Swart's convention,
the predictions for the hyperon couplings are
\begin{align}
g_{K\Lambda N}^{SU(3)}  &  =-\frac{g_{\pi NN}}{\sqrt{3}}\left(3 - 2\alpha
_{D}\right)  ,\label{su3l}\\
g_{K\Sigma N}^{SU(3)}  &  =g_{\pi NN}\left(  2\alpha_{D}-1\right)  ,
\label{su3s}%
\end{align}
where $\alpha_{D}=D/\left(  D+F\right)  $ is the standard fraction involving
$D$ and $F$-couplings. Using the $SU(6)$ value, $\alpha_{D}=0.6$, and
$g_{\pi NN}=13.02$, the
predicted $SU_{F}(3)$ hyperon coupling constants are $g_{K\Lambda N}%
^{SU(3)}=-13.53$ and $g_{K\Sigma N}^{SU(3)}=2.60$. However, flavor
symmetry is broken, typically quoted at least 20\% \cite{bsu3},
and we prefer using the experimental value $\alpha_{D}=0.644$
determined by Donoghue-Holstein \cite{Donoghue}. This yields the
couplings, $g_{K\Lambda N}^{SU(3)} = -12.87 $ and $ g_{K\Lambda
N}^{SU(3)} = 3.75$, and broken symmetry ranges, $-15.44 \leq
g_{K\Lambda N}^{SU(3)}\leq-10.30$ and $3.00 \leq g_{K\Sigma N}^{SU(3)}%
\leq 4.50 $.

We conclude this section by noting that there appears to be an
inconsistency in the literature regarding phases in the DW sum
rule, especially the relative sign between the nucleon and hyperon
discrepancies.  To ensure that Eq. (\ref{gapsum}) has the
appropriate phases, we use the $SU_F(3)$ relations, Eqs.
(\ref{su3l}) and (\ref{su3s}).  Although the sum rule does not
respect flavor symmetry, its derivation utilizes the $SU_F(3)$
representation for the current operator and thus the relative
signs between the GT discrepancies must be the same as given in
the $SU_F(3)$ limit.  Then eliminating $\alpha_D$ from Eqs.
(\ref{su3l}, \ref{su3s}) yields
\begin{equation}
-\frac{2 }{\sqrt{3}}g_{\pi NN} \ .
g_{K\Lambda N}^{SU(3)}    = \frac{g_{K\Sigma N}^{SU(3)} -2g_{\pi
NN}}{\sqrt{3}} \ .
\label{su3eq}
\end{equation}
Combining this result with the $SU_F(3)$ limit ($m_{K} \rightarrow
m_{\pi}$) of Eqs. (\ref{constraint}) and (\ref{b}) gives for the
GT couplings
\begin{equation}
(g^{GT}_{K\Lambda N})_{SU(3)}    = \frac{(g^{GT}_{K\Sigma N})_{SU(3)} -
2(g^{GT}_{\pi NN})_{SU(3)}}{\sqrt{3}} \ ,
\end{equation}
which has the same form (and signs) as Eq. (\ref{su3eq}).
Substituting the GT couplings from Eqs. (\ref{gngt}, \ref{glgt}, \ref{gsgt})
and taking the $SU_F(3)$ limit ($f_{\pi} =
f_{K}$,
$m_B = m_{B^\prime}$),  produces
\begin{equation}
(g^{\Lambda}_{A})_{SU(3)}    = \frac{(g^{\Sigma}_{A})_{SU(3)} -
2(g^{N}_{A})_{SU(3)}}{\sqrt{6}} \ .
\label{gtsu3}
\end{equation}
Finally, inserting the  $SU_F(3)$ axial charges, $(g^{N}_{A})_{SU(3)} = D +
F$,
$(g^{\Lambda}_{A})_{SU(3)} = -(D + 3 F)/\sqrt {6}$ and
$(g^{\Sigma}_{A})_{SU(3)} = D - F$,
Eq. ({\ref{gtsu3}) reduces to an identity verifying the phases and coefficients
in our sum rule are consistent.

\section{Application of the Criterion}

In the past decade there have been numerous  analyses of
hyperon reactions, most involving electromagnetic processes.  Table II
summarizes a large class
of phenomenological couplings  in  rationalized,
$g_{KYN}/\sqrt{4\pi}$, form. Several theoretical predictions are also included
for comparison. The results of Gobbi et al.~\cite{Scoccola} are based on a
Skyrme-type model, while Choe et al.~\cite{Choe} use a QCD sum rule method and
Jeong~\cite{Jeong} et al. employ the chiral bag model.
The broken $SU_{F}(3)$ and GT values are presented along
with the Taylor series extrapolation.

\begin{table}[h]
\caption{\label{tab:allcouplings}
Hyperon coupling constants.}
\begin{tabular}
[c]{||l|l|l||}\hline\hline
\textbf{Reference} & $g_{K\Lambda N}\mathbf{/}\sqrt{4\pi}$ & $g_{K\Sigma
N}/\sqrt{4\pi}$\\\hline\hline
\multicolumn{3}{||c||}{Phenomenological values}\\\hline
Adelseck et al. \cite{Adelseck} & $-4.17$ & $1.18$\\\hline
Williams et al. \cite{WJC 1} & $-1.16$ & $0.09$\\\hline
Williams et al. \cite{WJC 2} & $-2.38$ & $0.27$\\\hline
Mart et al. \cite{Mart} & $-1.84$ & $2.72$\\\hline
Mart et al. \cite{Mart} & $-0.84$ & $1.30$\\\hline
Mart et al. \cite{Mart} & $0.51$ & $0.13$\\\hline
David et al. \cite{David} & $-3.16$ & $0.91$\\\hline
David et al. \cite{David} & $-3.23$ & $0.80$\\\hline
Bennhold et al. \cite{Bennhold} & $-3.09$ & $1.23$\\\hline
Guidal et al. \cite{Guidal} & $-3.25$ & $1.26$\\\hline
Feuster et al. \cite{Feuster} & $-2.44$ to $-1.73$ & \\\hline
Lee et al. \cite{Lee} & $-3.80$ & $1.20$\\\hline
Hsiao et al. \cite{Hsiao} & $-2.41$ to $-1.24$ & $-0.50$ to $1.04$\\\hline
Chiang et al. \cite{Chiang} & $-2.38$ & $0.40$\\\hline
Janssen et al. \cite{Janssen 2} & $-0.4$ & \\\hline
Janssen et al. \cite{Janssen 3} & $-0.23$ & $0.28$\\\hline
Martin \cite{martin} & $-3.73^\dagger$ & $1.82^\dagger$\\\hline
Antolin \cite{ant} & $-3.53^\dagger$ & $1.53^\dagger$\\\hline
Timmermans et al. \cite{Timmermans} & $-3.86$ & $1.10$\\\hline\hline
\multicolumn{3}{||c||}{Theoretical values}\\\hline
Gobbi et al. \cite{Scoccola} & $-1.06$ & $0.37$\\\hline
Gobbi et al. \cite{Scoccola} & $-2.17$ & $0.76$\\\hline
Choe et al. \cite{Choe} & $-1.96$ & $0.33$\\\hline
Jeong  et al. \cite{Jeong} & $-3.77$ & $1.19$\\\hline
$SU_{F}(3)$\, $\alpha_D = .644$  & $-3.63$ & $1.06$\\\hline
$SU_{F}(3)$ broken at 20\% & $-4.\,\allowbreak4$ to $-2.\,\allowbreak9$ &
$0.8$ to $\allowbreak1.3$\\\hline
GT & $-2.60$ & $0.91$\\\hline
Taylor series & $-3.04$ & $1.34$\\\hline\hline
\end{tabular}
$^{\dagger} \text {Signs undetermined but taken the same as $SU_F(3)$}.$
\end{table}

In Figs. 2 and 3 we compare all tabulated $\Lambda$ and $\Sigma$ couplings,
respectively, in histogram form and also
indicate the broken $SU_{F}(3)$ interval (dark gray), GT   (diamond)
and Taylor series (inverted triangle) couplings.%

\begin{figure}[h]
\begin{center}
\includegraphics[trim=0.000000in 0.000000in 0.004941in
0.000000in,height=2.1067in,
width=2.559in]{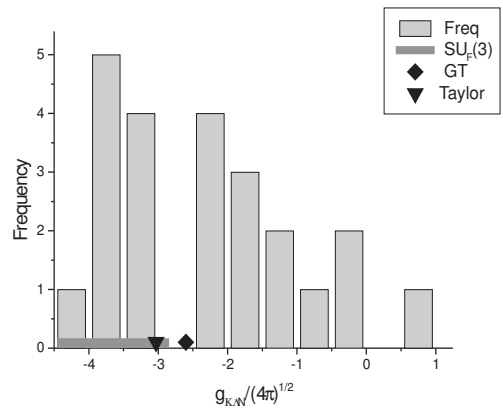}
\caption{GT, $SU_{F}(3)$,  Taylor series and published $K\Lambda N$
coupling constants.}
\end{center}%
\end{figure}
%

\begin{figure}[h]
\begin{center}
\includegraphics[
trim=0.000000in 0.000000in 0.000000in 0.241627in,height=1.8127in,width=2.501in]
{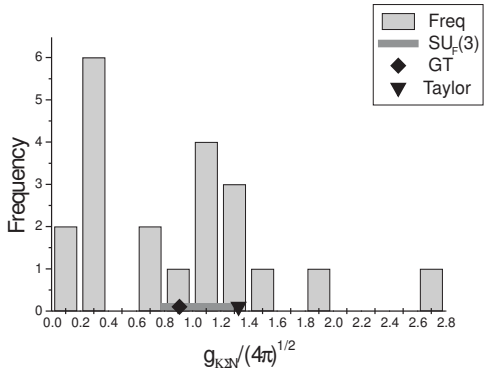}%
\caption{GT, $SU_{F}(3)$, Taylor series and published $K\Sigma N$ coupling
constants.}
\end{center}%
\end{figure}

Applying the constraint criterion,  the intercept $b$ requires
specifying $g_{\pi NN}$ which is still subject to discussion
\cite{Machleidt-Banerjee, de Swart 2,Uppsala 3, Pavan, Arndt,
Uppsala 2, Uppsala}. Nevertheless, it is generally accepted that
$12.90\leq g_{\pi NN}\leq13.20$. This uncertainty is the dominant
contribution to the variation in the intercept, $\Delta b$, given
by \bea \Delta b = \sqrt{\sum_i (\frac{\partial b} {\partial x_i}
\,\, \Delta x_i)^2} \ , \eea where the independent variables $x_i$
and the corresponding errors $\Delta x_i$ are the $\pi N$
coupling, the three axial charges $g_A^B$ and the two meson decay
constants $f_M$. Evaluating yields $b = -13.293$ and $\Delta b =
2.281$ which produces the maximum and minimum intercepts,
$b_{max}/\sqrt{4\pi}=-11.011/\sqrt{4\pi} = -3.106$ and
$b_{min}/\sqrt{4\pi}=-15.575/\sqrt{4\pi}= -4.394$. The
corresponding constraint lines are shown in Fig. 4.  A more
stringent constraint can be obtained from the value recommended by
de Swart et al. \cite{de Swart 2}, $g_{\pi NN}=13.02\pm.08$, which
produces the rationalized intercepts $-3.245$ and $-4.011$,
represented by the narrower band (dashed lines) in the figure.

Treating the coupling constants as coordinates, ($g_{K\Sigma N}, g_{K\Lambda
N}$), the values in Table II are  plotted in Fig. 4 (dark circles for
phenomenological, triangles for theoretical, box for $SU_{F}(3)$, diamond
for GT and inverted triangle for Taylor series). The couplings satisfying
the constraint are those with
coordinates within the band.
This produces the reduced or filtered set of acceptable
values  listed in Table III.
The histograms for this subset of $\Lambda$ and $\Sigma$ couplings are
depicted in Figs. 5 and 6,
respectively.
$\Lambda$ and


\begin{figure}[h]
\begin{center}
\includegraphics[height=2.2701in,width=3.0381in]{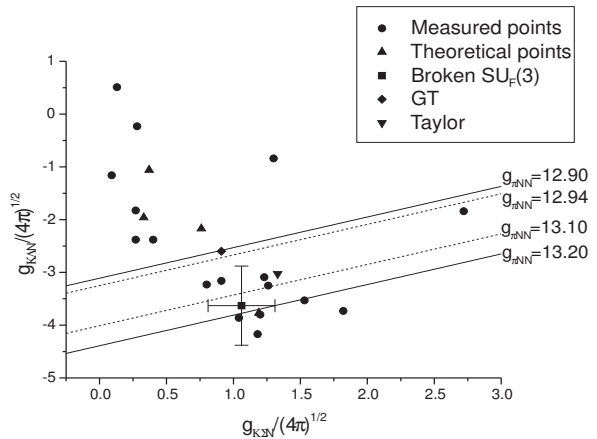}
\caption{Hyperon coupling coordinates and the constraint.}
\end{center}
\end{figure}


\begin{table}[h]
\caption{\label{tab:allcouplings}
Hyperon couplings consistent with the constraint.}
\begin{tabular}
[c]{||l|l|l||}\hline\hline
\textbf{Reference} & $g_{K\Lambda N}\mathbf{/}\sqrt{4\pi}$ & $g_{K\Sigma
N}/\sqrt{4\pi}$\\\hline\hline
\multicolumn{3}{||c||}{Phenomenological values}\\\hline
Mart et al. \cite{Mart} & $-1.84$ & $2.72$\\\hline
David et al. \cite{David} & $-3.16$ & $0.91$\\\hline
David et al. \cite{David} & $-3.23$ & $0.80$\\\hline
Bennhold et al. \cite{Bennhold} & $-3.09$ & $1.23$\\\hline
Guidal et al. \cite{Guidal} & $-3.25$ & $1.26$\\\hline
\multicolumn{3}{||c||}{Theoretical values}\\\hline
$SU_{F}(3)$ \, $\alpha_D = .644$  & $-3.63$ & $1.06$
\\\hline
GT   & $-2.60$ & $0.91$ \\\hline
Taylor series   & $-3.04$ & $1.34$ \\\hline\hline
\end{tabular}
\end{table}

\begin{figure}[h]
\begin{center}
\includegraphics[height=2.0859in,width=2.5088in]{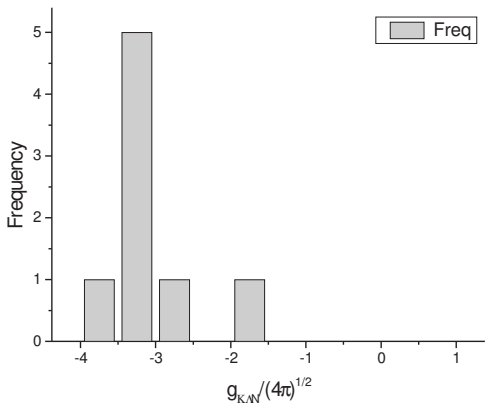}%
\caption{ Filtered $K\Lambda N$ coupling constants.}
\end{center}%
\end{figure}
%

\begin{figure}[h]
\begin{center}
\includegraphics[trim=0.000000in 0.000000in -0.048381in
0.105498in,height=2.0877in,
width=2.5097in]{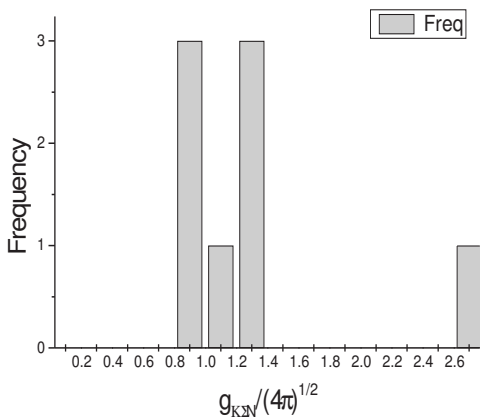}%
\caption{Filtered $K\Sigma N$ coupling constants.}
\end{center}%
\end{figure}

Because the criterion only applies to hyperon coupling pairs
$\left(  g_{K\Sigma N},g_{K\Lambda N}\right)$, two of the
$\Lambda$ analyses~\cite{Feuster,Janssen 2} were precluded from
Table III. This does not necessarily mean that their results are
unacceptable. Also note from Fig. 4 that not all of the broken
$SU_{F}(3)$ values are inside the constraint region and if the
more stringent constraint (dashed lines) is imposed, the $SU_F(3)$
centroid point
would be eliminated from
Table III along with the GT couplings. Clearly,
determining a more precise $g_{\pi N N}$ will be helpful in
obtaining accurate hyperon couplings.

The hyperon couplings consistent with the constraint fall in the
ranges $0.80\leq g_{K\Sigma N}/\sqrt{4\pi}\leq2.72$ and $-3.90\leq
g_{K\Lambda N}\mathbf{/}\sqrt{4\pi}\leq-1.84$.  The -3.90 limit on
the $\Lambda $ coupling was obtained from the intersection of the
constraint line with the area representing the broken $SU_F(3)$
uncertainty region.  Although we summarize our criterion analysis
by quoting these ranges, it is important to stress that the
constraint does not specify upper or lower bounds for the coupling
constants. However,  examining the filtered points it is
interesting that, with the exception of  one $\Lambda$ and one
$\Sigma$ value, all couplings have  magnitudes above the GT
predictions.  This suggests that the GT values may be lower
bounds, $\left\vert g_{KYN}\right\vert \geq\left\vert
g_{KYN}^{GT}\right\vert $, similar to the $\pi N$ coupling result,
$g_{\pi NN}\geq g_{\pi NN}^{GT}$. If this proves true then the
Taylor series extrapolations may be good estimates of the hyperon
coupling constants.

\section{Conclusion}

In summary, a hyperon coupling criterion has been developed using
the fundamental Goldberger-Treiman relation and the
Dashen-Weinstein sum rule. Because the $\pi N$ coupling constant
is not exactly known, the criterion can only restrict an area in
the ($g_{K\Sigma N}$, $g_{K\Lambda N}$) plane. As the precision of
$g_{\pi NN}$ improves, this area will decrease providing a
stronger constraint. Since the criterion can only be applied to
hyperon couplings in pairs, phenomenological investigations
incorporating this constraint should also perform a combined
$\Lambda$ and $\Sigma$ data analysis, especially since  the
couplings are  interrelated in most models. Even if there is model
dependence producing renormalized, effective hyperon couplings,
the  constraint should still be applicable to all hyperon
reactions.

This work has also applied the criterion to a large class of
published couplings to produce a reduced number of $g_{KYN}$
parameters satisfying the constraint. This subset spans the
intervals $0.80\leq g_{K\Sigma N}/\sqrt{4\pi}\leq2.72$ and
$-3.90\leq g_{K\Lambda N}\mathbf{/}\sqrt{4\pi}\leq -1.84$ and
includes the  $SU_{F}(3)$, GT and extrapolated Taylor series
couplings. The range of absolute values in the filtered sets
suggests that the GT coupling constants are, in magnitude, lower
bounds. Although further study is necessary to rigorously
demonstrate this, the constraint as a general guideline should be
reasonably useful provided the Dashen-Weinstein sum rule is
quantitatively valid.


\begin{acknowledgments}
The authors are grateful to T. Hare for informative comments.
This work was supported by the Department of Energy under grant
DE-FG02-97ER41048.
\end{acknowledgments}


\end{document}